\documentclass{JHEP3}\newcommand{\format} {\JHEPformat}
\usepackage{amsmath}
\usepackage{epsfig}
\usepackage{latexsym}

\newcommand{\JHEPformat} {
\bibliographystyle{JHEP}
\newcommand{\maketitlepage} {}
\abstract{\theabstract}
\keywords{\thekeywords}
\preprint{\thepreprint}
}

\newcommand{\TITLE}[1] {\newcommand{\thetitle} {#1}\title{#1}}
\newcommand{\ABSTRACT}[1] {\newcommand{\theabstract} {#1}}
\newcommand{\AUTHOR}[1] {\newcommand{\theauthor} {#1}}
\newcommand{\ADDRESS}[1] {\newcommand{\theaddress} {#1}}
\newcommand{\DATE}[1] {\newcommand{\thedate} {#1}\date{#1}}
\newcommand{\KEYWORDS}[1] {\newcommand{\thekeywords} {#1}}
\newcommand{\PREPRINT}[1] {\newcommand{\thepreprint} {#1}}

\newcommand{\p}{{\cal P}}
\newcommand{\N}{{\cal N}}
\newcommand{\HH}{{\cal H}}

\newcommand{\e}{{\epsilon}}

\newcommand{\ka}{{\kappa}}
\newcommand{\K}{{\rm K}}

\newcommand{\LL}{{\cal L}}
\newcommand{\tr} {\operatorname{tr}}
\newcommand{\sll} {\mathop{\rm sl}}

\newcommand{\bra}[1] {\left<#1\right|}
\newcommand{\ket}[1] {\left|#1\right>}
\newcommand{\braket}[2] {\left<#1\vphantom{#1}\right|
                         \left.\!\vphantom{#1}{#2}\right>}

\TITLE{Virasoro operators in the continuous basis of string field theory}

\AUTHOR{{\bf Ehud Fuchs}\footnote{\tt udif@tau.ac.il},
{\bf Michael Kroyter}\footnote{\tt mikroyt@tau.ac.il},
{\bf Alon Marcus}\footnote{\tt alon@tau.ac.il}}

\ADDRESS{School of Physics and Astronomy\\
  The Raymond and Beverly Sackler Faculty of Exact Sciences\\
  Tel Aviv University, Ramat Aviv, 69978, Israel
}

\author{Ehud Fuchs, Michael Kroyter, Alon Marcus\\
School of Physics and Astronomy\\
The Raymond and Beverly Sackler Faculty of Exact Sciences\\
Tel Aviv University, Ramat Aviv, 69978, Israel\\
E-mails:
\email{udif@tau.ac.il}, \email{mikroyt@tau.ac.il}, \email{alon@tau.ac.il}
}

\ABSTRACT{

In this work we derive two important tools for working in the $\ka$ basis 
of string field theory. First we give an analytical expression for the 
finite part of the spectral density $\rho_{\text{fin}}$. This expression 
is relevant when both matter and ghost sectors are considered.
Then we calculate the form of the matter part of the Virasoro generators 
$L_n$ in the $\ka$ basis, which construct string field theory's derivation 
$Q_{BRST}$. We find that the Virasoro generators are given by one dimensional 
delta functions with complex arguments.
}

\DATE{October 2002}
\PREPRINT{TAUP-2714-02\\{\tt hep-th/0210155}}

\format

\begin{document}

\maketitlepage

\section{Introduction and summary}

String field theory~\cite{Witten:1986cc} is a non-perturbative,
off-shell formulation of string theory.
The basic entities of this theory are fields over the space of string configurations.
These fields are multiplied using a non-commutative
star product, defined by the gluing of the right half of one
string to the left half of the other.
The derivation operator on these fields, that appears in the kinetic term,
is $Q_{BRST}$.

Gross and Jevicki~\cite{Gross:1987ia} gave an explicit formulation of the star
product in the (infinite dimensional) oscillator basis.
They defined the star product by an interaction vertex, which is a squeezed
state over a direct product of three Fock spaces.
This form of the vertex implies that squeezed states form a subalgebra of
the star product.

In~\cite{Rastelli:2001hh} Rastelli Sen and Zwiebach diagonalized these matrices
and found their spectrum.
The eigenvectors of these matrices form a continuous basis whose eigenvalues
are in the range $-\infty<\ka<\infty$.
Working in this basis simplifies calculations involving the star product.

Calculating normalization of squeezed states in the continuous basis
involves determinants of continuous matrices.
If these matrices are diagonal $X=X_\ka\delta(\ka{-}\ka')$,
then the determinant has the form
\begin{equation}
\det{X}=\exp\left( \tr\log X \right)=\exp\left( \int d\ka\delta(\ka-\ka)\log X_\ka \right).
\end{equation}
A similar expression holds for matrices in the $\HH_{\ka^2}$
subalgebra~\cite{Fuchs:2002zz}.
The delta function is the spectral density $\rho(\ka)$ of the continuous
basis.
This spectral density diverges, and in the level-truncation regularization
its divergence is $\ka$ independent and behaves as
$\rho^L(\ka)=\frac{\log L}{2\pi}$.
Because of this divergence, it seems that the determinant can only get
the values $0,1,\infty$
depending on whether the integral $\int d\ka\log X$ is negative,
zero or positive.
But $\rho(\ka)$ has a finite contribution $\rho_{\text{fin}}(\ka)$ which is $\ka$
dependent.
When the ghost sector contributions cancel these infinities, 
$\rho_{\text{fin}}(\ka)$ cannot be ignored, and the expression
$$
\exp \left( \int d\ka\rho_{\text{fin}}(\ka)\log X  \right)
\,,
$$
should be taken into account.
Belov and Konechny calculated $\rho_{\text{fin}}(\ka)$ numerically
in~\cite{Belov:2002pd}.
We find the analytic expression
\begin{equation}
\rho_{\text{fin}}(\ka)=
\frac{4\log(2)-2\gamma-\Psi(\frac{i \ka}{2})-\Psi(-\frac{i \ka}{2})}{4\pi}\,,
\end{equation}
where $\gamma$ is Euler's constant, and $\Psi$ is the digamma (polygamma)
function.

Another missing ingredient of the continuous basis
is the form of the Virasoro generators.
The Virasoro generators are used in string field theory to construct
$Q_{BRST}$, from which the kinetic term around the perturbative vacuum is built.
Other derivations built from the Virasoro generators can serve as kinetic
terms as well~\cite{Drukker:2002ct}.
The Virasoro generators
are also useful in the construction of surface states~\cite{LeClair:1989sp}, and
in particular of surface state projectors \cite{Gaiotto:2002kf}.
In~\cite{Douglas:2002jm} it was noticed that the expression for $L_0$ in
the continuous basis diverges. This is true for all the Virasoro generators.
Nonetheless, we manage to find an analytic expression for them.
For a single scalar field, $L_0$ is given by
\begin{equation}
\begin{aligned}
L_0&=\alpha' p_0^2+\sum_{n=1}^\infty n a^\dagger_n a_n=\alpha' p_0^2
+\int_{-\infty}^\infty
 \frac{d\ka\,d\ka'\,a^\dagger_{\ka'} a_{\ka}} {\sqrt{\N(\ka)\N(\ka')}}
  g_0^{\ka,\ka'}\,,\\
g_0^{\ka,\ka'}&\equiv\sum_{n=1}^\infty n v_n^{\ka}v_n^{\ka'}
=\frac{\cosh\bigl(\frac{(\ka+\ka') \pi}{4}\bigr)}{2}
\left(\delta(\ka{-}\ka'{+}2i)+\delta(\ka{-}\ka'{-}2i)\right).
\end{aligned}
\end{equation}
The use of complex arguments in the one dimensional delta function is
somewhat unorthodox,
but the definition of the delta function for complex arguments is essentially
the same as for real arguments.
We elaborate on the definition of the delta function and demonstrate its
use in the body of the paper.

The complex delta functions hide the high divergence of $L_0$
in the continuous basis. They appear in all the Virasoro generators.
The midpoint preserving reparameterization generators
$K_n=L_n-(-1)^n L_{-n}$ should have milder divergences~\cite{Erler:2002nr},
and indeed they do not contain complex delta functions.

This paper is organized as follows.
In section~\ref{review} we fix our notations for the continuous basis
and give a useful integral expression for the eigenvectors $v_n^{\ka}$.
In section~\ref{rhoFin} we calculate the finite part of the spectral density.
In section~\ref{prem} we introduce a useful operator, which we denote 
$\LL_\ka$, and elaborate on the definition and use of the one dimensional
delta function with a complex argument.
We use these tools in section~\ref{Virasoro} to calculate the
Virasoro generators. Sections~\ref{prem} and~\ref{Virasoro} are independent 
of section~\ref{rhoFin}.

\section{The continuous basis}
\label{review}

In this section we summarize the properties of the continuous basis,
and set our notations.
In~\cite{Rastelli:2001hh} the star algebra derivation $K_1=L_1+L_{-1}$ was
represented by a matrix over the oscillators basis
$[K_1,v\cdot a]=(\K_1 v)\cdot a$
and its spectrum was found to be
\begin{equation}
\K_1 v^{\ka}=\ka v^{\ka}\,,
\end{equation}
where $-\infty<\ka<\infty$ and $v^\ka$ are the eigenvectors of the matrix 
$\K_1$. The generating function of these eigenvectors is
\begin{equation}
\label{fkaz}
f_\ka(z)\equiv\sum_{n=1}^\infty \frac{v^{\ka}_n}{\sqrt{n}}z^n
=\frac{1}{\ka}\left(  1-e^{-\ka\tan^{-1}z}\right).
\end{equation}
This relation can be inverted to give
\begin{equation}
\label{contDef}
v_n^{\ka}=\frac{\sqrt{n}}{2\pi i}\oint \frac {f_\ka(z)}{z^{n+1}}dz\,.
\end{equation}

Okuyama~\cite{Okuyama:2002yr} found the orthogonality and completeness
relations of the eigenvectors
\begin{equation}
\label{Comp}
\sum_{n=1}^\infty v_n^{\ka}v_n^{\ka'}=\N(\ka)\delta(\ka-\ka')\,,\qquad
\int_{-\infty}^\infty \frac{d\ka}{\N(\ka)} v_n^{\ka}v_m^{\ka}=\delta_{n,m}\,,
\end{equation}
where the normalization is given by
\begin{equation}
\N(\ka)=\frac{2}{\ka}\sinh\left(\frac{\ka\pi}{2}\right).
\end{equation}
The continuous basis creation and annihilation operators are given by
\begin{equation}
\label{aka2an}
a^\dagger_\ka=\sum_{n=1}^\infty \frac{v_n^{\ka}}{\sqrt{\N(\ka)}}\, a^\dagger_n
\,, \qquad
a_\ka=\sum_{n=1}^\infty  \frac{v_n^{\ka}}{\sqrt{\N(\ka)}}\, a_n\,,
\end{equation}
where $a_n$ are the usual modes of the string. The inverse relations are
\begin{equation}
\label{an2aka}
a^\dagger_n=\int_{-\infty}^\infty d\ka \frac{v_n^{\ka}}{\sqrt{\N(\ka)}}
  \,a^\dagger_\ka\,, \qquad
a_n=\int_{-\infty}^\infty d\ka \frac{v_n^{\ka}}{\sqrt{\N(\ka)}}\, a_\ka\,.
\end{equation}
With these relations the $\ka$ oscillators satisfy the canonical commutation 
relations
\begin{equation}
\label{akakdCom}
[a_\ka,a^\dagger_{\ka'}]=\delta(\ka{-}\ka')\,.
\end{equation}

We end this section by giving
an integral representation for $v^{\ka}_n$~\cite{Fuchs:2002zz} that is more
useful then~(\ref{contDef}) in some calculations.
\begin{equation}
\begin{split}
v_{2n-1}^{\ka}&=
 \frac{(-1)^{n+1}\sqrt{2n-1}}{\pi}\N(\ka)\int_1^\infty
   \frac{\cos(\kappa \coth^{-1}(x))}{x^{2n}}dx\\
&=\frac{(-1)^{n+1}\sqrt{2n-1}}{\pi}\N(\ka)\int_0^\infty
   \frac{\cos(\ka u)\tanh^{2n-2}(u)}{\cosh^{2}(u)}du\,,
\end{split}
\label{v2n-1Exp}
\end{equation}
where we substituted $x=\coth(u)$ in the last step.
The analogous expression for $v_{2n}$ reads
\begin{equation}
\begin{aligned}
v_{2n}^{\ka}&=
\frac{(-1)^{n}\sqrt{2n}}{\pi}\N(\ka)\int_1^\infty
 \frac{\sin(\kappa \coth^{-1}(x))}{x^{2n+1}}dx\\
&=\frac{(-1)^{n}\sqrt{2n}}{\pi}\N(\ka)\int_0^\infty
 \frac{\sin(\ka u)\tanh^{2n-1}(u)}{\cosh^{2}(u)}du\,.
\end{aligned}
\label{v2nExp}
\end{equation}
Collectively they can be written as
\begin{equation}
\label{colExp}
v_n^{\ka}=
\frac{i^{n-1}\sqrt{n}}
{2\pi}\N(\ka)\int_{-\infty}^\infty
\frac{e^{i \kappa u}\tanh^{n-1}(u)}{\cosh^2(u)}du\,.
\end{equation}
These expressions will allow us to change the order of summation and 
integration in the calculations of the spectral density and the Virasoro 
generators.

\section{The spectral density}
\label{rhoFin}

The spectral density $\rho(\ka)$ is needed for calculating 
traces and determinants of operators that are diagonal in the continuous 
basis.
Ignoring for a moment the fact that the trace is base invariant,
we can calculate
\begin{equation}
\text{tr}G=\bra{n}G\ket{n}=\braket{n}{\ka}\bra{\ka}G\ket{\ka'}\braket{\ka'}{n}
\,.
\end{equation}
Now, using the fact that $G$ is diagonal in the $\ka$ basis
\begin{equation}
\bra{\ka}G\ket{\ka'}=G_\ka\delta(\ka{-}\ka')\,,
\end{equation}
we get
\begin{equation}
\text{tr}G=\braket{n}{\ka}\braket{\ka}{n}G_\ka
  =\rho(\ka)G_\ka\,,
\end{equation}
where we define the spectral density
\begin{equation}
\rho(\ka)\equiv\braket{\ka}{n}\braket{n}{\ka}
   =\frac{1}{\N(\ka)}\sum_{n=1}^\infty v_n^\ka v_n^\ka\,.
\end{equation}
From the invariance of the trace $\text{tr}G=\bra{\ka}G\ket{\ka}$, it is 
obvious that $\rho(\ka)$ diverges like $\delta(0)$. In~\cite{Rastelli:2001hh} 
it was shown that the leading term in level truncation regularization 
is $\rho^L(\ka)=\frac{\log L}{2\pi}$, where $L$ denotes the level.
This term is $\ka$ independent. In~\cite{Belov:2002pd} the finite, $\ka$ 
dependent, contribution to the spectral density was defined as
\begin{equation}
\label{rhoFinDef}
\rho_{\text{fin}}^{2L}(\ka)=\frac{1}{\N(\ka)}\sum_{n=1}^{2L}v_n^{\ka}v_n^{\ka}-
\frac{1}{2\pi}\sum_{n=1}^{L}\frac{1}{n}\,,
\end{equation}
and evaluated numerically.
In this section we obtain an analytical expression for this term
\begin{equation}
\rho_{\text{fin}}(\ka)=\lim_{L\rightarrow \infty}\rho_{\text{fin}}^{2L}(\ka)\,.
\end{equation}

We begin by regularizing the two diverging sums on the r.h.s 
of~(\ref{rhoFinDef}) with powers of the variable $z$ to get
\begin{equation}
\label{rhoDef}
\rho_{\text{fin}}(z,\ka)=\frac{1}{\N(\ka)}\sum_{n=1}^{\infty}\left(
  v_{2n-1}^{\ka}v_{2n-1}^{\ka}z^n + v_{2n}^{\ka}v_{2n}^{\ka}z^n\right) -
\frac{1}{2\pi}\sum_{n=1}^{\infty}\frac{z^n}{n}\,,
\end{equation}
where $|z|<1$. The last sum gives
\begin{equation}
-\frac{1}{2\pi}\sum_{n=1}^{\infty}\frac{z^n}{n}=\frac{1}{2\pi}\log(1-z)\,.
\end{equation}

We evaluate the first sum by using eq.~(\ref{v2n-1Exp})
\begin{equation}
\begin{split}
\label{vvOdd}
\frac{1}{\N(\ka)}&\sum_{n=1}^{\infty}v_{2n-1}^{\ka}v_{2n-1}^{\ka}z^n=
    \frac{2z^\frac{3}{2}}{\N(\ka)}\partial_z\sum_{n=1}^{\infty}
             \frac{v_{2n-1}^{\ka}v_{2n-1}^{\ka}}{2n-1}z^{n-\frac{1}{2}} \\
&=\frac{2\N(\ka) z^\frac{3}{2}}{\pi^2}\partial_z\int_0^\infty  du\,dv
\frac{\sqrt{z}\cos(\ka u) \cos(\ka v)}{\cosh^2(u)\cosh^2(v)}
               \sum_{n=1}^\infty (z\tanh^2(u)\tanh^2(v))^{n-1} \\
&=\frac{2\N(\ka) z^\frac{3}{2}}{\pi^2}
     \partial_z\int_0^\infty   \frac{\sqrt{z}\cos(\ka u) \cos(\ka v) }
         {\cosh^2(u)\cosh^2(v)-z \sinh^2(u)\sinh^2(v)} du\,dv \\
&=\frac{z}{\pi}
     \int_0^\infty   \frac{\cos(\ka v) \cos(\ka \tanh^{-1}(\sqrt{z}\tanh(v))}
         {\cosh^2(v)-z\sinh^2(v)} dv \, .
\end{split}
\end{equation}
In the last step we replaced $\int_0^\infty du \cos(\ka u)...$ with
$\frac{1}{2}\int_{-\infty}^\infty du \exp(i \ka u)...$ and
evaluated the $u$ integral by
closing the contour in the upper half plane for $\ka>0$ (with analytic 
continuation for $\ka\leq 0$) picking up the residues at
$u=\frac{(2n-1)\pi i}{2}\pm \tanh^{-1}(\sqrt{z}\tanh(v))$.

In a similar way we evaluate the second term on the r.h.s of~(\ref{rhoDef})
\begin{equation}
\begin{aligned}
\label{vvEven}
\frac{1}{\N(\ka)} \sum_{n=1}^{\infty}&v_{2n}^{\ka}v_{2n}^{\ka}z^n=
    \frac{2z}{\N(\ka)}\partial_z\sum_{n=1}^{\infty}
             \frac{v_{2n}^{\ka}v_{2n}^{\ka}}{2n}z^{n}\\
&=\frac{2}{\pi^2}\N(\ka)
   z\partial_z\int_0^\infty   \frac{z\sin(\ka u) \sin(\ka v) \tanh(u)\tanh(v)}
         {\cosh^2(u)\cosh^2(v)-z \sinh^2(u)\sinh^2(v)} du\,dv\\
&=\frac{\sqrt{z}}{\pi}
     \int_0^\infty   \frac{\sin(\ka v) \sin(\ka \tanh^{-1}(\sqrt{z}\tanh(v))}
         {\cosh^2(v)-z\sinh^2(v)} dv \,,
\end{aligned}
\end{equation}
where now, in addition 
to the residues at $u=\frac{(2n-1)\pi i}{2}\pm \tanh^{-1}(\sqrt{z}\tanh(v))$, 
there are also residues at $u=\frac{(2n-1)\pi i}{2}$. 

In the $z\rightarrow 1$ limit we add the integrals in equations~(\ref{vvOdd}) 
and~(\ref{vvEven}) and neglect terms that behave like $(1-z)\log(1-z)$ to get
\begin{equation}
\rho_{\text{fin}}(\ka)=\lim_{z\rightarrow1}
\frac{1}{\pi}
     \int_0^\infty \frac  {\cos\left(\ka\left(v- \tanh^{-1}(\sqrt{z}\tanh(v))\right)\right)}
         {\cosh^2(v)-z\sinh^2(v)} dv+\frac{1}{2\pi}\log(1-z)\,.
\end{equation}
To evaluate this integral we write $\sqrt{z}=1-\epsilon$, and drop terms which would
become irrelevant in the limit
\begin{equation}
\begin{aligned}
\label{rhoInt}
\rho_{\text{fin}}(\ka)= & \lim_{\epsilon\rightarrow 0}\left(
\frac{1}{\pi}
     \int_0^\infty \frac  {\cos\left(\frac{\ka}{2}\log(1+\frac{\epsilon}{2} 
e^{2v})\right)}
         {1+\frac{\epsilon}{2}e^{2v}} dv+\frac{1}{2\pi}\log(2\epsilon)\right)\\= &
\lim_{\epsilon\rightarrow 0} \frac{1}{2\pi}\left(
   \int_{\frac{\epsilon}{2}}^\infty \frac{\cos\bigl(\frac{\ka \xi}{2}\bigr)}{e^\xi-1}d\xi
            +\log(2\epsilon)\right),
\end{aligned}
\end{equation}
where $\xi=\log(1+\frac{\epsilon}{2} e^{2v})$.
To get rid of the infinities in the limit we calculate
\begin{equation}
\partial_\ka\rho_{\text{fin}}(\ka)=
-\frac{1}{4\pi}
   \int_0^\infty \frac{\xi\sin\bigl(\frac{\ka \xi}{2}\bigr)}{e^\xi-1}d\xi\,,
\end{equation}
and since $\rho_{\text{fin}}(0)=\frac{\log(2)}{\pi}$
(as can be seen from eq.~(\ref{rhoInt}) or from a direct calculation)
we get
\begin{equation}
\rho_{\text{fin}}(\ka)=\frac{\log(2)}{\pi}+
       \int_0^\ka(\partial_\ka\rho_{\text{fin}}(\ka)) d\ka=
\frac{4\log(2)-2\gamma-\Psi(\frac{i \ka}{2})-\Psi(-\frac{i \ka}{2})}{4\pi}\,,
\end{equation}
where $\gamma$ is Euler's constant, and $\Psi$ is the digamma (polygamma) function.
This result agrees with the numerical estimates of~\cite{Belov:2002pd}.

\section{Mathematical preliminary}
\label{prem}

In the next section we shall see that in order to calculate the Virasoro
generators in the $\ka$ basis, we need to introduce the differential operator 
$\LL_\ka$. This operator is a sum of two shift operators in the imaginary 
direction. We define and explain the use of this operator in~\ref{LLDefSec}.

We shall have to operate with $\LL_\ka$ on delta functions, generating 
delta functions with complex arguments, creatures that we usually do 
not encounter.
Thus, in~\ref{compDelta} we elaborate on the definition of these delta 
functions and demonstrate their use.

\subsection{The operator $\LL_\ka$}
\label{LLDefSec}

Our task in this section is to find an operator $\LL_\ka$ which obeys
\begin{equation}
\label{LLDef}
\LL_\ka (\ka v_n^{\ka})=n v_n^{\ka}\,.
\end{equation}
Equivalently this operator should obey the relation
\begin{equation}
\LL_\ka \frac{\ka v_n^{\ka}}{\sqrt{n}}z^n=z \partial_z \frac{v_n^{\ka}}{\sqrt{n}}z^n.
\end{equation}
Summing over $n$, and using the generating function~(\ref{fkaz}),
we get
\begin{equation}
\label{LLEq}
\LL_\ka (1-e^{-\ka u})=\frac{1}{2}\sin(2u)e^{-\ka u}\,,
\end{equation}
where $u=\tan(z)$.
We see that
\begin{equation}
\label{LLSol}
\LL_\ka=\frac{1}{2}\sin(2\partial_\ka)
\end{equation}
satisfies the above equation.

The operator $\LL_\ka$ is a sum of two shift operators
\begin{equation}
\label{LLfDef}
\LL_\ka f(\ka)=\frac{\exp(2i\partial_\ka)-\exp(-2i\partial_\ka)}{4i}f(\ka)=
                     \frac{f(\ka+2i)-f(\ka-2i)}{4i}\,,
\end{equation}
and as such is defined only for functions which can be Taylor expanded with convergence radius
$r>2$.
Since $\LL_\ka$ contains only odd powers of $\partial_\ka$ we get for
functions with convergence radius $r>2$ on the real axis
an integration by parts formula
\begin{equation}
\label{IntByParts}
\int_{-\infty}^\infty f(\ka)\LL_\ka g(\ka)d\ka=
    -\int_{-\infty}^\infty g(\ka)\LL_\ka f(\ka)d\ka\,.
\end{equation}

The case $r=2$ should be handled with care. We illustrate it with an example
that will be used in section~\ref{Virasoro}. 
\begin{equation}
\exp(2i\partial_\ka)\frac{1}{\cosh\bigl(\frac{(\ka-\ka') \pi}{4}\bigr)}=
   \frac{1}{\cosh\bigl(\frac{(\ka-\ka'+2i) \pi}{4}\bigr)}=
   \frac{-i}{\sinh\bigl(\frac{(\ka-\ka') \pi}{4}\bigr)}\,,
\end{equation}
where $\ka'\neq\ka$.
For $\ka'=\ka$ the radius of convergence is $r=2$,
and the above expression is undefined. To extract the singular part of it we write
\begin{equation}
\begin{aligned}
\label{singular}
\exp(2i\partial_\ka)\frac{1}{\cosh\bigl(\frac{(\ka-\ka') \pi}{4}\bigr)}
  &=\lim_{\e\rightarrow 0}\left(
  \exp((2-\e)i\partial_\ka)\frac{1}{\cosh\bigl(\frac{(\ka-\ka') \pi}{4}\bigr)}
   \right)\\=
\lim_{\e\rightarrow 0}
   \frac{-i}{\sinh\bigl(\frac{(\ka-\ka') \pi}{4}\bigr)-i \e}& =
 -i \p \frac{1}{\sinh\bigl(\frac{(\ka-\ka')\pi}{4}\bigr)}+
      \pi \delta\left(\sinh\Bigl(\frac{(\ka-\ka')\pi}{4}\Bigr) \right)\\&=
 -i \p \frac{1}{\sinh\bigl(\frac{(\ka-\ka')\pi}{4}\bigr)}+4\delta(\ka{-}\ka')
 \,,
\end{aligned}
\end{equation}
where $\p$ represents the principal value of the function. In a similar fashion
\begin{equation}
\exp(-2i\partial_\ka)\frac{1}{\cosh\bigl(\frac{(\ka-\ka') \pi}{4}\bigr)}=
 i \p \frac{1}{\sinh\bigl(\frac{(\ka-\ka') \pi}{4}\bigr)}+4 \delta(\ka{-}\ka') \,.
\end{equation}
We can take the symmetric and antisymmetric parts of these
shift operators and write
\begin{equation}
\begin{aligned}
\label{coshSA}
\cos(2\partial_\ka)\left(\frac{1}{\cosh\bigl(\frac{(\ka-\ka') \pi}{4}\bigr)}
  \right)&= 4\delta(\ka{-}\ka')\,,\\
\sin(2\partial_\ka)\left(\frac{1}{\cosh\bigl(\frac{(\ka-\ka') \pi}{4}\bigr)}
  \right)&= -\p \frac{1}{\sinh\bigl(\frac{(\ka-\ka') \pi}{4}\bigr)}\,.
\end{aligned}
\end{equation}
The antisymmetric part is of importance because of~(\ref{LLSol}).
The importance of the symmetric part lies in the ``trigonometric'' identity
\begin{equation}
\label{TrigId}
\sin(2\partial_\ka)(f(\ka)g(\ka))=(\sin(2\partial_\ka)f(\ka)) (\cos(2\partial_\ka)g(\ka))+
       (\cos(2\partial_\ka)f(\ka)) (\sin(2\partial_\ka)g(\ka))\,.
\end{equation}
We shall expand the definition of $\LL_\ka$ to functions
whose radius of convergence is $r<2$ at the end of the next subsection.

\subsection{The one dimensional delta function with complex argument}
\label{compDelta}

In this subsection we present the properties of the complex delta function,
as it will emerge when we shall operate with $\LL_\ka$ on $\delta(\ka{-}\ka')$ in
the calculations of the Virasoro generators.
Formally, $\LL_\ka\delta(\ka{-}\ka')$ is
an infinite sum of delta function derivatives,
but does this sum converge to a distribution? We demonstrate that this sum is
an object very similar to a regular distribution.

So far, we defined the operation of $\LL_\ka$ only on functions with 
convergence radius $r>2$. However, for the delta function, there is no 
notion of convergence radius because it is a distribution. We consider the 
definition of the delta function as a ``limit'' of functions,
\begin{equation}
\label{deltaLimit}
\delta(\ka)=\lim_{\e\rightarrow 0}\delta_\e(\ka)\,.
\end{equation}
Different $\delta_\e$ sequences have different radii of convergence. While the
sequence used in~(\ref{singular}) has a zero radius, other sequences, 
such as the limit of gaussians,
\begin{equation}
\delta_\e(\ka)=\frac{1}{\sqrt{\pi\e}}e^{-\frac{\ka^2}{\e}}\,,
\end{equation}
are analytic in the whole 
complex plane.
Henceforth, we define the delta function using a sequence of this type.
We can now define
\begin{equation}
\LL_\ka\delta(\ka{-}\ka')=\lim_{\epsilon\rightarrow 0}\LL_\ka
\delta_\epsilon(\ka{-}\ka')\equiv
\frac{1}{4i}(\delta(\ka{-}\ka'{+}2i)-\delta(\ka{-}\ka'{+}2i))\,.
\end{equation}
The complex arguments of the delta functions may seem strange, but in 
fact integrations involving these delta functions are similar to the 
familiar case of a real argument. This can be seen using the behaviors 
at (real directed) infinity of $\delta_\epsilon$ and contour 
arguments\footnote{Note that $z\neq 0$ does not imply $\delta(z)=0$.
In fact getting expressions such as $\delta(2i)$ in a result would
probably mean that a non-legitimate manipulation was performed on the way.
Also note, that while $\delta_\e(\ka{-}\ka'{\pm}2i)$ are complex, the 
combination $\LL_\ka\delta_\epsilon(\ka{-}\ka')$ is a real function.}.
These contour arguments only apply when we convolute the delta function 
with an analytic function $f(\ka)$ that has no poles on the way to the new 
contour of integration. Thus, the complex delta function acts as a 
distribution, when confined to this class of functions.

Suppose now that $f(\ka)$ has simple poles $\ka_n$ in the 
range $0\leq\Im(\ka)\leq 2$. To evaluate
$\int_{-\infty}^\infty\delta(\ka{-}2i)f(\ka)d\ka$,
we displace the contour to the line $\Im(\ka)=2$, and pick up the
residues along the way. Poles for which $0<\Im(\ka_n)<2$, contribute
$2\pi i\delta(\ka_n{-}2i) \mbox{res}_{\ka_n}(f(\ka))$, while poles for 
which $\Im(\ka_n)=2$ contribute 
$\pi i\delta(\ka_n{-}2i) \mbox{res}_{\ka_n}(f(\ka))$. The last case
to consider is the case of integrating the principle part of poles 
located on the real line. In this case we again get a contribution 
of $\pi i\delta(\ka_n{-}2i) \mbox{res}_{\ka_n}(f(\ka))$ to the integral. All 
in all
\begin{equation}
\begin{aligned}
\label{resDelta+}
\int_{-\infty}^\infty  &\delta(\ka{-}2i)\p f(\ka)d\ka\\&=f(2i)+
  2\pi i  \!\!\!\sum_{0<\Im(\ka_n)<2}\!\!\!\mbox{res}_{\ka_n}(f(\ka))
\delta(\ka_n{-}2i)+
   \pi i \!\!\!\sum_{\Im(\ka_n)=0,2}\!\!\!\mbox{res}_{\ka_n}(f(\ka))
\delta(\ka_n{-}2i)\,.
\end{aligned}
\end{equation}
Due to a change in the orientation of integration
\begin{equation}
\begin{aligned}
\label{resDelta-}
\int_{-\infty}^\infty  &\delta(\ka{+}2i)\p f(\ka)d\ka\\&=f(-2i)-
  2\pi i  \!\!\! \!\!\!\sum_{-2<\Im(\ka_n)<0\!\!\!}\!\!\!\mbox{res}_{\ka_n}(f(\ka))
\delta(\ka_n{+}2i)-
   \pi i \!\!\! \!\!\!\sum_{\Im(\ka_n)=-2,0}\!\!\!\!\!\!\mbox{res}_{\ka_n}(f(\ka))
\delta(\ka_n{+}2i)\,.
\end{aligned}
\end{equation}
We see that unlike regular distributions, the convolution of these 
generalized distributions with analytic functions can result in generalized
distributions, rather then functions. The incorporation of multiple poles 
is straightforward and adds terms with derivatives of the delta function, 
but we shall not need it here.

Finally we introduce a recipe for handling expressions such as 
$\LL_\ka f(\ka)$, where $f(\ka)$ has radius of convergence $r<2$
\begin{equation}
\label{LLDeltaDef}
\LL_\ka f(\ka)= \LL_\ka \int \delta(\ka-\tilde \ka)f(\tilde \ka) d\tilde\ka
   \equiv \int  \LL_\ka \delta(\ka-\tilde \ka)f(\tilde \ka) d\tilde\ka \,.
\end{equation}
This definition involves changing the order of integration with the
limit~(\ref{deltaLimit}).
As we naturally think of this limit as being taken after all integrations 
were performed, we shall refer to the r.h.s of this equation as the 
definition of $\LL_\ka f(\ka)$. For the case $r>2$ this definition coincides 
with~(\ref{LLfDef}).

\section{The Virasoro generators in the $\ka$ basis}
\label{Virasoro}

In this section we obtain the form of the Virasoro generators in the $\ka$ 
basis for a single scalar field.
Since the expressions are cumbersome, we start in~\ref{l0} by finding $L_0$, 
which is the simplest one. It is also the most useful one, in
particular when working in the Siegel gauge. In subsection~\ref{lpm1}
we calculate $L_{\pm 1}$ and confirm the closure of the $\sll(2)$ algebra 
directly. Next, we get the general expression for all the generators and 
obtain identities that follow from the Virasoro algebra. All of the above 
is done in the zero momentum sector for clarity.
We conclude by giving the general expressions for the 
generators including the zero mode.

\subsection{Calculating $L_0$}
\label{l0}

Formally, writing the Virasoro generator $L_0$ in the $\ka$ basis involves
a simple change of basis
\begin{equation}
L_0=\sum_{n=1}^\infty n a^\dagger_n a_n=
   \sum_{n=1}^\infty \int_{-\infty}^\infty \frac{d\ka\,d\ka'\,
    a^\dagger_{\ka'} a_\ka n  v_n^{\ka}v_n^{\ka'}}
       {\sqrt{\N(\ka)\N(\ka')}} =\int_{-\infty}^\infty
 \frac{d\ka\,d\ka'\,a^\dagger_{\ka'} a_\ka} {\sqrt{\N(\ka)\N(\ka')}}
    g_0^{\ka,\ka'}\,,
\end{equation}
where we define\footnote{
The matrix $g_0^{\ka,\ka'}$ will play the role of a contravariant metric 
in~\cite{FKM2}.}
\begin{equation}
\label{gdef}
g_0^{\ka,\ka'}\equiv\sum_{n=1}^\infty n v_n^{\ka}v_n^{\ka'}\,.
\end{equation}
However, it was shown in~\cite{Douglas:2002jm} that this sum is highly 
divergent. Thus, $g_0^{\ka,\ka'}$ cannot be given by a regular function.
We derive an analytic expression for it as a generalized distribution by 
acting on both sides of the completeness relation~(\ref{Comp}) with 
the $\LL_\ka$ operator
\begin{equation}
\begin{aligned}
\label{g^kkEq}
g_0^{\ka,\ka'}&=\LL_\ka (\ka\sum_{n=1}^\infty v_n^{\ka}v_n^{\ka'})
=\LL_\ka (\ka \N(\ka)\delta(\ka{-}\ka'))\\
&=\frac{\sinh\bigl(\frac{\ka' \pi}{2}\bigr)}{2i}
    (\delta(\ka{-}\ka'{+}2i)-\delta(\ka{-}\ka'{-}2i))\\
&=\frac{\cosh\bigl(\frac{(\ka'+\ka)\pi}{4}\bigr)}{2}(\delta(\ka{-}\ka'{+}2i)
+\delta(\ka{-}\ka'{-}2i))
\,,
\end{aligned}
\end{equation}
where in the last step we used the properties of the delta function 
in order to give a manifestly symmetric expression for $g_0^{\ka,\ka'}$.

We conclude this subsection with a consistency check.
Define the covariant metric by the following convergent sum
\begin{equation}
\label{gIdef}
g^0_{\ka,\ka'}\equiv\sum_{n=1}^\infty \frac{v_n^{\ka}v_n^{\ka'}}{n}\,.
\end{equation}
By the completeness relation~(\ref{Comp}) and the definition of
$\LL_\ka$~(\ref{LLDef}), $g^0_{\ka,\ka'}$ obeys the identity
\begin{eqnarray}
\label{Iofg}
\LL_\ka (\ka g^0_{\ka,\ka'})=\N(\ka)\delta(\ka{-}\ka')\,.
\end{eqnarray}
We evaluate this metric by the methods 
of~\cite{Okuyama:2002yr} and obtain 
\begin{equation}
g^0_{\ka,\ka'}=
\frac{2\sinh\bigl(\frac{\ka \pi}{4}\bigr)
    \sinh\bigl(\frac{\ka' \pi}{4}\bigr)}
    {\ka \ka' \cosh\bigl(\frac{(\ka-\ka') \pi}{4}\bigr)}\,.
\end{equation}
This expression can be used to verify eq.~(\ref{Iofg}) directly
\begin{equation}
\begin{aligned}
\label{deltaFromSech}
\LL_\ka (\ka g^0_{\ka,\ka'})
  &=\frac{\sinh\bigl(\frac{\ka' \pi}{4}\bigr)}{\ka'}
       \left(\sin(2\partial_\ka)\sinh\left(\frac{\ka \pi}{4}\right)\right)
        \left(\cos(2\partial_\ka)\frac{1}
                   {\cosh\bigl(\frac{(\ka-\ka') \pi}{4}\bigr)}\right) \\
  &=\frac{4\sinh\bigl(\frac{\ka' \pi}{4}\bigr)
      \cosh\left(\frac{\ka \pi}{4}\right)\delta(\ka{-}\ka')}{\ka'}=
    \N(\ka)\delta(\ka{-}\ka')\,,
\end{aligned}
\end{equation}
where we used eq.~(\ref{coshSA}),(\ref{TrigId}), and the fact that
\begin{equation}
\cos(2\partial_\ka)\sinh\left(\frac{\ka \pi}{4}\right)=0\,, \qquad
\sin(2\partial_\ka)\sinh\left(\frac{\ka \pi}{4}\right)
    =\cosh\left(\frac{\ka \pi}{4}\right).
\end{equation}
Using the complex delta function integration 
rules~(\ref{resDelta+}),(\ref{resDelta-}), one can 
verify that the metrics obey
\begin{equation}
\label{gIsgInv}
\int_{-\infty}^\infty \frac{d\tilde\ka}{\N(\tilde\ka)}
    g^0_{\ka,\tilde\ka}g_0^{\tilde\ka,\ka'}=
    \N(\ka)\delta(\ka{-}\ka')\,,
\end{equation}
as expected.

\subsection{Calculating $L_{\pm 1}$}
\label{lpm1}

We use similar methods to obtain the remaining $\sll(2)$ generators.
Actually, it is enough to find $L_{-1}$, since $L_1$ can be
obtained by hermitian conjugation. We write $L_{-1}$ in the $\ka$ basis
\begin{equation}
\begin{aligned}
L_{-1}&=\sum_{n=1}^\infty \sqrt{n(n+1)} a^\dagger_{n+1} a_n \\
  &=\sum_{n=1}^\infty \int_{-\infty}^\infty \frac{d\ka\,d\ka'\,
                a^\dagger_{\ka'} a_\ka}
        {\sqrt{\N(\ka)\N(\ka')}}
 \sqrt{n(n+1)} v_n^{\ka} v_{n+1}^{\ka'}=
         \int_{-\infty}^\infty \frac{d\ka\,d\ka'\,a^\dagger_{\ka'} a_\ka}
                  {\sqrt{\N(\ka)\N(\ka')}} g^{\ka,\ka'}_1\,,
\end{aligned}
\end{equation}
where
\begin{equation}
g^{\ka,\ka'}_1 \equiv \sum_{n=1}^\infty \sqrt{n(n+1)} v_n^{\ka} 
v_{n+1}^{\ka'}\,.
\end{equation}

Since the sum does not converge, we define the well behaved 
expression\footnote{Note that unlike 
$g_0^{\ka,\ka'},g^0_{\ka,\ka'}$ which play 
the role of metrics in suitable spaces, $g^{\ka,\ka'}_1,g_{\ka,\ka'}^1$ 
have no such interpretation. In fact they are not even symmetric with 
respect to $\ka,\ka'$.}
\begin{equation}
\label{gkaka1}
g_{\ka,\ka'}^1\equiv\sum_{n=1}^\infty \frac{1}{\sqrt{n(n+1)}} v_n^{\ka} 
v_{n+1}^{\ka'}\,.
\end{equation}
We will get $g^{\ka,\ka'}_1$ using
\begin{equation}
\label{g1ofg1}
g^{\ka,\ka'}_1=\LL_{\ka'}\LL_\ka (\ka \ka' g_{\ka,\ka'}^1)\,.
\end{equation}
Substituting the integral expression for $v_n^\ka$~(\ref{colExp}) we obtain
\begin{equation}
\begin{aligned}
\label{g_1Final}
g_{\ka,\ka'}^1 &=i\frac{\N(\ka)\N(\ka')}
   {4\pi^2} \int_{-\infty}^\infty
 \frac{du\,dv\,e^{i (\ka u+\ka' v)}}{\cosh^2(u) \cosh^2(v)}
     \sum_{n=1}^\infty (-1)^{n-1} \tanh^{n-1}(u) \tanh^n(v)\\&=
i\frac{\N(\ka)\N(\ka')}
   {4 \pi^2} \int_{-\infty}^\infty
 \frac{du\,dv\,e^{i (\ka u+\ka' v)}\tanh(v)}{\cosh(u)\cosh(v)\cosh(u+v)}\\&=
\frac{1}{\ka}-\frac{(\ka-\ka')\sinh\bigl(\frac{\ka' \pi}{2}\bigr)}
                 {2\ka \ka' \sinh\bigl(\frac{(\ka-\ka') \pi}{4}\bigr)
          \cosh\bigl(\frac{(\ka-\ka') \pi}{4}\bigr)}\,.
\end{aligned}
\end{equation}
A direct calculation using~(\ref{coshSA}),(\ref{TrigId}) gives
\begin{equation}
\label{BadLuck}
\LL_\ka (\ka g_{\ka,\ka'}^1)=\p\frac{\sinh\bigl(\frac{\ka' \pi}{2}\bigr)}
    {\ka' \sinh\bigl(\frac{(\ka-\ka') \pi}{2}\bigr)}\,.
\end{equation}
As this expression has zero radius of convergence for $\ka=\ka'$,
we use~(\ref{LLDeltaDef}) and get
\begin{equation}
\begin{aligned}
g^{\ka,\ka'}_1&=\LL_{\ka'}\int_{-\infty}^\infty d\tilde\ka\,
       \delta(\tilde\ka{-}\ka')
       \p\frac{\sinh\bigl(\frac{\tilde\ka \pi}{2}\bigr)}
  {\sinh\bigl(\frac{(\ka-\tilde\ka) \pi}{2}\bigr)}\\
&=\int_{-\infty}^\infty d\tilde\ka\, 
    \p\frac{(\delta(\tilde\ka{-}\ka'{-}2i)-\delta(\tilde\ka{-}\ka'{+}2i))
        \sinh\bigl(\frac{\tilde\ka \pi}{2}\bigr)}
  {4i \sinh\bigl(\frac{(\ka-\tilde\ka) \pi}{2}\bigr)}\\
&=-\frac{\sinh\bigl(\frac{\ka \pi}{2}\bigr)}{2}
   (\delta(\ka{-}\ka'{+}2i)+\delta(\ka{-}\ka'{-}2i)+2\delta(\ka{-}\ka'))\,,
\end{aligned}
\end{equation}
where in the last step we used the integration 
rules~(\ref{resDelta+}),(\ref{resDelta-}).

To verify that the $\sll(2)$ algebra indeed holds, we write
\begin{equation}
\begin{aligned}[]
 [L_1,L_{-1}]&=\int_{-\infty}^\infty \frac{d\ka_1\,d\ka_2\,d\ka_3\,d\ka_4\,
  \sinh\left(\frac{\ka_1 \pi}{2}\right)
  \sinh\left(\frac{\ka_3 \pi}{2}\right)}{4\sqrt{\N(\ka_1)\N(\ka_2)\N(\ka_3)\N(\ka_4)}}
     [a^\dagger_{\ka_1}a_{\ka_2},a^\dagger_{\ka_4}a_{\ka_3}]\\
 &\qquad\quad\times(\delta(\ka_1{-}\ka_2{+}2i)+\delta(\ka_1{-}\ka_2{-}2i)+2\delta(\ka_1{-}\ka_2))\\
 &\qquad\quad\times(\delta(\ka_3{-}\ka_4{+}2i)+\delta(\ka_3{-}\ka_4{-}2i)+2\delta(\ka_3{-}\ka_4))\\
&=
\int_{-\infty}^\infty \frac{d\ka\,d\ka'\,d\tilde\ka\,a^\dagger_{\ka'}a_\ka}
    {4\sqrt{\N(\ka)\N(\ka')}}
\frac{\sinh\left(\frac{\ka \pi}{2}\right)\sinh\bigl(\frac{\ka' \pi}{2}\bigr)-
        \sinh^2\left(\frac{\tilde\ka \pi}{2}\right)}{\N(\tilde\ka)}\\
 &\qquad\quad\times(\delta(\ka{-}\tilde\ka{+}2i)+\delta(\ka{-}\tilde\ka{-}2i)+2\delta(\ka{-}\tilde\ka))\\
 &\qquad\quad\times(\delta(\ka'{-}\tilde\ka{+}2i)+\delta(\ka'{-}\tilde\ka{-}2i)+2\delta(\ka'{-}\tilde\ka))=2L_0\,.
\end{aligned}
\end{equation}
The derivation of the other commutation relations is similar.

\subsection{General $L_n$}
\label{ln}

Our strategy should be familiar by now, with a minor change, the appearance
of creation (annihilation) operators bilinears. We write
\begin{equation}
\begin{aligned}
\label{Lcont}
L_{-m}&= \frac{1}{2}\sum_{n=1}^{m-1}\sqrt{n(m-n)}a^\dagger_n a^\dagger_{m-n}+
       \sum_{n=1}^\infty \sqrt{n(n+m)} a^\dagger_{n+m} a_n \\
&= \frac{1}{2}\int_{-\infty}^\infty \frac{d\ka\,d\ka'\,a^\dagger_{\ka'}a^\dagger_\ka}
        {\sqrt{\N(\ka)\N(\ka')}} h^{\ka,\ka'}_m+
         \int_{-\infty}^\infty \frac{d \ka d \ka' a^\dagger_{\ka'} a_\ka}
                  {\sqrt{\N(\ka)\N(\ka')}} g^{\ka,\ka'}_m\,,\\
L_m = L_{-m}^\dagger &=
     \frac{1}{2}\int_{-\infty}^\infty \frac{d\ka\,d\ka'\,a_{\ka'}a_\ka}
        {\sqrt{\N(\ka)\N(\ka')}} h^{\ka,\ka'}_m+
         \int_{-\infty}^\infty \frac{d \ka d \ka' a^\dagger_{\ka'} a_\ka}
                  {\sqrt{\N(\ka)\N(\ka')}} g^{\ka',\ka}_m\,,
\end{aligned}
\end{equation}
where
\begin{equation}
\begin{aligned}
g^{\ka,\ka'}_m& \equiv \sum_{n=1}^\infty \sqrt{n(n+m)} v_n^{\ka} 
v_{n+m}^{\ka'}\,,\\
h^{\ka,\ka'}_m& \equiv \sum_{n=1}^{m-1} \sqrt{n(m-n)} v_n^{\ka} 
v_{m-n}^{\ka'}\,.
\end{aligned}
\end{equation}

We start from calculating
\begin{align}
\label{g_m}
g_{\ka,\ka'}^m&\equiv\sum_{n=1}^\infty \frac{1}{\sqrt{n(n+m)}}
                 v_n^{\ka} v_{n+m}^{\ka'}\nonumber\\
  &=i^m\frac{\N(\ka)\N(\ka')}
    {4\pi^2}\int_{-\infty}^\infty
 \frac{du\,dv\,e^{i (\ka u+\ka' v)}}{\cosh^2(u) \cosh^2(v)}
     \sum_{n=1}^\infty (-1)^{n-1} \tanh^{n-1}(u) \tanh^{n+m-1}(v)\nonumber\\
  &=i^m\frac{\N(\ka)\N(\ka')}
    {4\pi^2} \int_{-\infty}^\infty
 \frac{du\,dv\,e^{i (\ka u+\ka' v)}\tanh^m(v)}{\cosh(u)\cosh(v)\cosh(u+v)}\\
  &=i^{m+1} \frac{\N(\ka')}
   {2\pi \ka} \int_{-\infty}^\infty
 dv \frac{\tanh^{m-1}(v)}{\cosh^2(v)}\left(e^{i (\ka'-\ka) v}-e^{i \ka' v}
\right)\nonumber .
\end{align}
We can now write
\begin{equation}
\begin{aligned}
\label{gm}
g^{\ka,\ka'}_m= \LL_{\ka'}\LL_\ka(\ka \ka' g_{\ka,\ka'}^m)&=
\LL_{\ka'}\LL_\ka \frac{i^{m+1} \ka'\N(\ka')}{2\pi} \int_{-\infty}^\infty
 dv \frac{\tanh^{m-1}(v)}{\cosh^2(v)}e^{i (\ka'-\ka) v} \\&=
\ka'\N(\ka ')\LL_{\ka'}\LL_\ka\frac{v_m^{\ka'-\ka}}{\sqrt{m}\N(\ka'-\ka)}\,,
\end{aligned}
\end{equation}
where in the second equality we noticed that the $\ka$ independent term
drops under the action of the derivative. In the last equality 
we recognized the integral expression for $v_n^\ka$~(\ref{colExp}),
and used the anticommutation relation
\begin{equation}
\left\{\LL_{\ka'},\ka'\N(\ka ')\right\}=0\,.
\end{equation}

In order to continue we recall that $v_m^{\ka'-\ka}$ are polynomials 
with respect to $\ka,\ka'$, and thus have infinite radius of convergence. 
Next, we notice that 
$\LL_{\ka'}\LL_{\ka}$ acts on a function of
$\ka_- \equiv \ka'-\ka$. We define $\ka_+ \equiv \ka+\ka'$, and get
\begin{align}
g^{\ka,\ka'}_m&=
   \frac{\sinh\bigl(\frac{\ka' \pi}{2}\bigr)}{8}
     (\cos(4\partial_-)-\cos(4\partial_+)) \frac{\ka_- v_m^{\ka_-}}
   {\sqrt{m}\sinh\left(\frac{\ka_- \pi}{2}\right)}\\
  &=\frac{\sinh\bigl(\frac{\ka' \pi}{2}\bigr)}{16}
     \int_{-\infty}^\infty  d\tilde\ka\,
    (\delta(\ka_-{-}\tilde\ka{+}4i)+\delta(\ka_-{-}\tilde\ka{-}4i)-
           2\delta(\ka_-{-}\tilde\ka))
      \frac{\tilde\ka v_m^{\tilde\ka}}
   {\sqrt{m}\sinh\left(\frac{\tilde\ka \pi}{2}\right)}\nonumber\\
\nonumber
  &=\sinh\Bigl(\frac{\ka' \pi}{2}\Bigr)
\biggl(
\frac{q_m(\ka_-)}{\sinh\left(\frac{\ka_- \pi}{2}\right)}+
\frac{i^m \delta(\ka_-{-}2i)-(-i)^m\delta(\ka_-{+}2i)}{2i} 
   -m\sin\Bigl(\frac{m\pi}{2}\Bigr)\delta(\ka_-)   \biggr)\,,
\end{align}
where we used
\begin{equation}
\frac{v_m^{\pm 2i}}{\sqrt{m}}=(\mp i)^{m-1}\,,\qquad
      \frac{v_m^{\pm 4i}}{\sqrt{m}}=(\mp i)^{m-1}m\,,
\end{equation}
and defined the polynomial
\begin{equation}
q_m(\ka)\equiv\frac{(\ka +4i)v_m^{\ka +4i}+(\ka -4i)v_m^{\ka -4i}-2\ka 
v_m^{\ka}}{16\sqrt{m}}=\oint 
\frac{e^{-\ka \tan^{-1}(z)}dz}{z^{m-1}(1+z^2)^2}\,.
\end{equation}

Finally, we find $h^{\ka,\ka'}_m$
\begin{align}
\label{hAlt}
\nonumber
&h^{\ka,\ka'}_m =\LL_{\ka'}\LL_\ka \left(\ka\ka'\sum_{n=1}^{m-1} 
\frac{1}{(2\pi i)^2 \ka \ka'}\oint \frac{dz\,dw}{z^{n+1}w^{m-n+1}}
e^{-\ka \tan^{-1}(z)-\ka' \tan^{-1}(w)}\right)\\
 &=\LL_{\ka'}\LL_\ka
  \oint  \frac{dz\,dw}{(2\pi i)^2}\left(\frac{1}{w^{m+1}}\left(\frac{1}{z-w}-\frac{1}{z}\right)+
      \frac{1}{z^{m+1}}\left(\frac{1}{w-z}-\frac{1}{w}\right)\right)
         e^{-\ka \tan^{-1}(z)-\ka' \tan^{-1}(w)}\nonumber\\&=
-\LL_{\ka'}\LL_\ka\frac{\ka_+ v_m^{\ka_+}}{\sqrt{m}}=q_m(\ka_+)\,.
\end{align}

The Virasoro generators of a single scalar field obey the $c=1$ Virasoro 
algebra
\begin{equation}
\label{VirAlg}
[L_n,L_m]=(n-m)L_{n+m}+\frac{n^3-n}{12}\delta_{n+m} \,.
\end{equation}

Substituting the form of the Virasoro generators in the continuous
basis~(\ref{Lcont}) in the algebra~(\ref{VirAlg}) gives the identities
\begin{subequations}
\begin{alignat}{2}
\label{ghConda}
(n-m)g_{n+m} &=\left[g_n,g_m \right]\,, \\
(n-m)h_{n+m} &=h_n g_m+(h_n g_m)^{\rm T}-h_m g_n-(h_m g_n)^{\rm T}\,, \\
\label{ghCondc}
\mbox{Tr}(h_m h_n)&=\frac{1}{6}\delta_{n,m}(n^3-n)\,, \\
(n+m)g_{n-m} &=g_n g_m^{\rm T}- g_m^{\rm T}g_n+h_m h_n && n\ge m\,, \\
(n+m)h_{n-m}&= g_m h_n+(g_m h_n)^{\rm T} && n\ge m\,, \\
0&=g_n h_m+(g_n h_m)^{\rm T} && n\ge m\,.
\end{alignat}
\end{subequations}
In these expressions matrix multiplication is assumed, and the summation of a
continuous index $\tilde \ka$, in the
multiplication as well as in the trace, should be understood as integration
$\int\frac{d\tilde \ka}{\N(\tilde \ka)}$. 

To demonstrate the formalism we give a direct proof of~(\ref{ghConda})
and~(\ref{ghCondc}). In computing the commutators of $g_m$ we use its
form in eq.~(\ref{gm}) and the expression for $v_m^\ka$ in~(\ref{colExp})
\begin{align}
\left[g_n,g_m \right]^{\ka\ka '} &=\ka '\N(\ka ')\LL_\ka\LL_{\ka'}
\int\frac{d\tilde \ka}{\sqrt{nm}}
\tilde \ka\LL_{\tilde \ka}\left(\frac{v_n^{\tilde \ka-\ka}}{\N(\ka-\tilde \ka)}
\right)
\LL_{\tilde \ka}\left(\frac{v_m^{\ka'-\tilde \ka}}{\N(\tilde \ka-\ka ')}
\right)-(n\leftrightarrow m)
\nonumber\\
&= \ka '\N(\ka ')\LL_\ka\LL_{\ka'}\LL_\ka\LL_{\ka'}
\int\frac{d\tilde \ka}{\sqrt{nm}}
\tilde \ka\frac{v_n^{\tilde \ka-\ka}v_m^{\ka'-\tilde \ka}}{\N(\ka-\tilde \ka)
\N(\tilde \ka-\ka ')}-(n\leftrightarrow m)
\nonumber\\
&=\ka '\N(\ka ')\LL_\ka^2\LL_{\ka'}^2\frac{i^{n+m-1}}{2\pi}\int du
\partial_u\left(\frac{e^{-i\ka u}\tanh^{n-1}u}{\cos^2{u}}\right)
\frac{e^{i\ka 'u}\tanh^{m-1}u}{\cos^2{u}}-(n\leftrightarrow m)
\nonumber\\
&=(n-m)\ka '\N(\ka ')\LL_\ka\LL_{\ka'}
\frac{v_{n+m}^{\ka '-\ka}}{\sqrt{n+m}\N(\ka-\ka ')}=(n-m)g_{n+m}^{\ka\ka'}\,,
\end{align}
where in the second step we used the symmetry property 
$\LL_{\tilde \ka} f(\ka-\tilde \ka)=-\LL_\ka f(\ka-\tilde \ka)$
and in the third step we performed 
the integration with respect to $\tilde \ka$. In the last step
we acted with one of the $\LL_\ka$ operators and with one $\LL_{\ka'}$.

For the identity~(\ref{ghCondc}) we use generating function techniques.
The r.h.s of the identity gives
\begin{equation}
\sum_{n,m\ge 1}z^mw^n\frac{1}{6}\delta_{n,m}(n^3-n)=\frac{z^2w^2}{(1-zw)^4}\,,
\end{equation}
while for the l.h.s
\begin{equation}
\begin{aligned}
\sum_{n,m\ge 1} z^m w^n &\int\frac{h_n^{\ka\ka '}h_m^{\ka '\ka}}{\N(\ka)\N(\ka ')}
d\ka\,d\ka ' \\
  &=\frac{z^2 w^2}{(1+z^2)^2 (1+w^2)^2}\int\frac{d\ka\,d\ka'}{\N(\ka)\N(\ka ')}
e^{-(\ka+\ka ')(\tan^{-1}z+\tan^{-1}w)} \\
  &=\frac{z^2 w^2}{(1+z^2)^2 (1+w^2)^2}
\left(\frac{1}{\cos^2(\tanh^{-1}z+\tanh^{-1}w)}\right)^2
=\frac{z^2w^2}{(1-zw)^4}\,,
\end{aligned}
\end{equation}
where we used~(\ref{hAlt}) to sum over $n,m$.

\subsection{The zero mode}

Adding the zero mode to the Virasoro generators is simple, as all we have 
to do is to substitute
\begin{equation}
\begin{aligned}
L_{-n}&\rightarrow L_{-n}+\sqrt{n}a^\dagger_n \sqrt{2\alpha'}
      p_0 =L_{-n}+\sqrt{2\alpha'}\int d\ka 
  \frac{\sqrt{n}v_n^\ka a^\dagger_\ka}{\sqrt{\N(\ka)}}p_0\,,\\
L_{n}&\rightarrow L_{n}+\sqrt{n}a_n  \sqrt{2\alpha'}p_0 =L_{n}+\sqrt{2\alpha'}\int d\ka 
  \frac{\sqrt{n}v_n^\ka a_\ka}{\sqrt{\N(\ka)}} p_0\,,\\
L_{0}&\rightarrow L_{0}+\alpha' p_0^2\,.
\end{aligned}
\end{equation}
There are new identities that follow from the Virasoro algebra~(\ref{VirAlg})
\begin{subequations}
\begin{alignat}{2}
(n-m)(\sqrt{n+m}v_{n+m})&=g_m^{\rm T} (\sqrt{n}v_n)-g_n^{\rm T} 
(\sqrt{m}v_m)\,,\\
0 &=g_n (\sqrt{m}v_m)+h_m (\sqrt{n}v_n) && \qquad n\ge m \,,\\
(n+m)(\sqrt{n-m}v_{n-m}) &=g_m (\sqrt{n}v_n)+h_n (\sqrt{m}v_m) && 
\qquad n\ge m\,.
\end{alignat}
\end{subequations}
This completes the derivation of the matter sector Virasoro generators.

\section{Conclusions}

In this paper we found the finite part of the spectral density
and the form of the Virasoro generators in the
continuous basis.
We hope that it would help solve some of the withstanding problems of string
field theory.

The most pretentious goal is to find an analytic solution to string field
theory's equation of motion $Q_{BRST}\Psi+\Psi*\Psi=0$.
One way to achieve this might be to follow the analytical methods
of~\cite{Kostelecky:2000hz} in the continuous basis.

It would also be interesting to attempt to calculate string amplitudes
analytically.
An expression for general string amplitudes was given recently
in~\cite{Taylor:2002bq}.
The advantage of this expression is that it involves determinants of
infinite matrices, making the level truncation calculations much simpler.
The matrices involved in this calculation are the Neumann coefficients,
coming from the vertices, and the Virasoro operators, coming from the
propagators.
Using the spectroscopy of the Neumann coefficients, together with our
results for the Virasoro operators and the spectral density in the
continuous basis, could allow for analytic calculations of string amplitudes.

For performing these tasks some more work is in order. Regarding 
$\rho_{\text{fin}}(\ka)$ one still has to understand how to incorporate 
it in the equations, as there is a discrepancy between CFT results 
and calculations involving $\rho_{\text{fin}}(\ka)$~\cite{Belov:2002pd}.
The analytical expression for the spectral density implies that this is not 
a numerical artifact.
As for the Virasoro generators, the form of the ghost sector is needed.
The expressions in the ghost sector resemble those of the matter
sector~\cite{Erler:2002nr}. Thus, they can be calculated
using the methods of this paper.

\acknowledgments

We would like to thank Jacob Sonnenschein, Yaron Oz and Ofer Aharony for 
fruitful discussions.
This work was supported in part by the US-Israel Binational Science
Foundation, the German-Israeli Foundation for Scientific Research,
and the Israel Science Foundation.

\bibliography{FKM}

\end{document}